\documentclass[aps,prb,twocolumn,showpacs]{revtex4}
\usepackage{bm}
\usepackage{graphicx}
\usepackage{amsmath}
\usepackage{eufrak}
\usepackage{color}
\usepackage{ulem}
\newcommand{\nix}[1]{}
\usepackage{graphics}

\usepackage{dcolumn}
\DeclareMathAlphabet{\mathitb}{OT1}{cmr}{bx}{sl}

\begin{document}

\title{Circular photogalvanic effect in HgTe/CdHgTe quantum well structures}
\author{B.~Wittmann,$^1$ S.N.~Danilov,$^1$
 V.V.~Bel'kov,$^{2}$ S.A.~Tarasenko,$^2$ E.G.~Novik,$^3$
H.~Buhmann,$^3$ C.~Br\"{u}ne,$^3$ L.W.~Molenkamp,$^3$
Z.D.~Kvon,$^4$ N.N.~Mikhailov,$^4$ S.A.~Dvoretsky,$^4$
N.\,Q.\,Vinh,$^5$ A.\,F.\,G.~van~der~Meer,$^5$ B.~Murdin,$^6$ and
S.D.~Ganichev$^{1}$ }
\affiliation{$^1$Terahertz Center, University of Regensburg, 93040 Regensburg, Germany}
\affiliation{$^2$A.F.\,Ioffe Physical-Technical Institute of the Russian Academy of Sciences, 194021 St.\,Petersburg, Russia}
\affiliation{$^3$ Physical Institute (EP3), University of
W\"{u}rzburg, 97074 W\"{u}rzburg, Germany}
\affiliation{$^4$ Institute of Semiconductor Physics, 630900
Novosibirsk, Russia}
\affiliation{$^5$ FOM Institute for Plasma Physics ``Rijnhuizen'', P.O.
Box 1207, NL-3430 BE Nieuwegein, The Netherlands}
\affiliation{$^6$University of Surrey, Guildford, GU2 7XH, UK}

\begin{abstract}
We describe the observation of the circular and linear
photogalvanic effects in HgTe/CdHgTe quantum wells. The interband
absorption of  mid-infrared radiation as well as the intrasubband
absorption of terahertz (THz) radiation in the QWs structures is
shown to cause a \textit{dc} electric current due to these
effects. The photocurrent magnitude and direction varies with the
radiation polarization state and crystallographic orientation of
the substrate in a simple way that can be understood from a
phenomenological theory. The observed dependences of the
photocurrent on the radiation wavelength and temperature
 are discussed.
\end{abstract}
\pacs{73.21.Fg, 72.25.Fe, 78.67.De, 73.63.Hs}

\maketitle

\section{Introduction}

High mobility HgTe quantum well (QW) structures are attracting
rapidly growing attention due to the peculiar band structure
properties. HgTe quantum wells exhibit an inverted band structure
ordering for QWs exceeding a certain critical width, i.e., the
ordering of electron- and hole-like states is
interchanged.~\cite{Buhmann09,Koenig07} In this kind of structures
the structural inversion asymmetry leads to a strong Rashba
spin-orbit splitting with energies of the order of 30
meV.~\cite{Gui04} Furthermore the quantum spin Hall
effect~\cite{Koenig07} has been observed in the bulk insulating
regime and manifests itself in the formation of two spin polarized
counter propagating one dimensional helical edge channels which
results in a quantized conductance without magnetic field. The
latter demonstrates that such inverted HgTe structures are an
example for a topologically non-trivial
insulator.~\cite{Bernevig06} Inverted band structure and strong
spin-orbit interaction also give rise to unusual opto-electronic
phenomena, e.g., a recently observed nonlinear magneto-gyrotropic
photogalvanic effect.~\cite{Diehl09} Photogalvanic effects (PGE),
in particular the circular photogalvanic effect (CPGE), in low
dimensional structures have been proved be very powerful for the
study of non-equilibrium processes in semiconductor QWs yielding
information on their point-group symmetry, inversion asymmetry of
QW structures, details of the band spin-splitting, processes of
electron momentum, spin, and energy relaxation (for a recent
review see Ref.~\onlinecite{IvchenkoGanichev_book}). The CPGE is a
photon helicity-dependent photocurrent, caused by a transformation
of the photon angular momenta into a translational motion of
charge carriers. Microscopically, the conversion can be due to
spin-dependent
mechanisms~\cite{Ganichev01,GanichevPrettlreview,Bieler05} or
orbital effects.~\cite{Tarasenko07,Olbrich09} The mechanisms may,
as shown theoretically in Refs.~\onlinecite{Deyo09,Moore09},
contain contributions associated with the Berry curvature and side
jumps giving rise to a macroscopic $dc$ current quadratic in the
amplitude of the $ac$ electric field.

Here we report the observation and investigation of the circular
photogalvanic effect in HgTe QW structures with different growth
direction induced by terahertz (THz) as well as mid-infrared
radiation. Our investigations show that the CPGE can be
effectively generated in HgTe quantum wells with a strength of
about an order of magnitude larger than that observed in GaAs,
InAs and SiGe low dimensional
structures.~\cite{IvchenkoGanichev_book} This large photocurrent
magnitude is of particular importance for an optimization of
all-electric semiconductor room temperature detectors, which are
based on photogalvanic effects and, which provide information
about the polarization state of the laser
radiation.~\cite{JAP2008} We present the phenomenological theory
of the linear photogalvanic effect (LPGE) and CPGE in
two-dimensional channels with point-group symmetry corresponding
to (001)- and (013)-oriented structures and compare the results
with experimental data on polarization dependences. We show the
results of the band structure calculations and discuss possible
mechanisms of infrared/THz radiation absorption.

\section{Samples and experimental technique}
\label{samples}

\begin{figure}[t]
\includegraphics[width=0.99\linewidth]{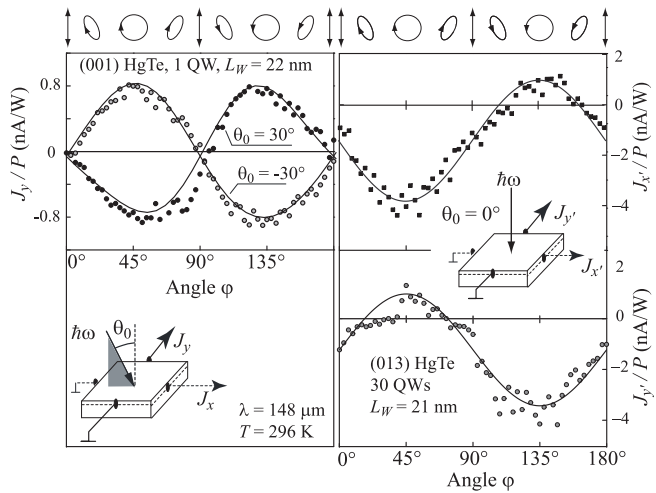}
\caption{Photocurrent as a function of radiation helicity measured
at room temperature in HgTe QWs grown on (001)- and (013)-oriented
substrates (left and right panels, respectively). The photocurrent
was excited by THz radiation with wavelength
$\lambda$~=~148~$\mu$m and power $P \approx 5$~kW. The left panel
shows a photocurrent generated in (001)-grown QW at oblique
incidence in the direction normal to the plane of incidence (data
are shown for $\theta_0 = \pm 30^\circ$).  The solid lines are
fits with the phenomenological equation $J_y = \pm J_0 \cdot \sin2
\varphi \propto P_{\rm circ}$. The right panels represent the
photocurrent in (013)-grown QWs detected at normal incidence
($\theta_0 =0$) in two perpendicular directions. Insets show
corresponding experimental geometries. Along the top the
polarization ellipses corresponding to key phase angles $\varphi$
are sketched. } \label{fig01}
\end{figure}

The experiments were carried out on $n$-type
Hg$_{0.3}$Cd$_{0.7}$Te/HgTe/Hg$_{0.3}$Cd$_{0.7}$Te quantum well
structures of (001) and (013) crystallographic orientation and
various quantum well widths $L_W$. Structures with (001) surface
orientation were grown by molecular beam epitaxy (MBE) on a
Cd$_{0.96}$Zn$_{0.04}$Te substrate and have $L_W$ either 8~nm or
22~nm. Structures with (013) surface orientation were grown on a
GaAs substrate by a modified MBE method~\cite{MBE} and have $L_W$
= 21~nm. Samples with sheet density of electrons $n_s$ from $1
\times 10^{11}$~cm$^{-2}$ to $2 \times 10^{12}$~cm$^{-2}$ and
mobility at 4.2~K in the range between $5 \times 10^4$ and $5
\times 10^5$ cm$^2$/Vs at $T=4.2$~K were studied in the
temperature range from 4.2\,K up to room temperature. The samples
were closely to square shaped of size $5 \times 5$~mm$^2$. While
for (001) QWs all edges were obtained by cleaving, for (013)
samples two edges were prepared by cleaving and the others were
cut normal to them. We  fabricated two pairs of ohmic contacts
prepared by thermal In-bonding and centered in the middle of
sample edges (see insets in Fig.~\ref{fig01}). In this way we
enabled investigations of photocurrents for (001)-grown QWs in $x
\parallel [1 \bar{1} 0]$ and $y \parallel [1 1 0]$
crystallographic directions and in (013) oriented samples in $x^\prime$ and $y^\prime$
directions. The geometry of the experiment is sketched in the insets in
Fig.~\ref{fig01}. The photocurrent was measured in unbiased
structures via the voltage drop across a $50\: \Omega$ load
resistor. The voltage was recorded with a storage oscilloscope.
For low-temperature measurements the samples were mounted in  an optical cryostat
with $z$-cut crystal quartz windows which allowed us to study photocurrents
in the temperature range from 4.2~K up to room temperature.

The photocurrents were induced  by direct band-to-band optical
transitions or indirect intrasubband (Drude-like) optical
transitions within the lowest size-quantized subband, applying
mid-infrared (MIR) or terahertz (THz) radiation, respectively. We
note, that in some samples transitions between size-quantized
subbands may contribute to the absorption of MIR radiation. As a
radiation source in the MIR range between 6~$\mu$m and 15~$\mu$m
(corresponding to photon energies from $\hbar \omega = 206.7$~meV
to 82.7~meV) we used a frequency tunable free electron laser
``FELIX'' at FOM-Rijnhuizen in the
Netherlands.~\cite{Knippels99p1578} The output pulses of light
from FELIX were chosen to be $\approx$6 ps long, separated by
40~ns, in a train (or ``macropulse'') of 7~$\mu$s duration. The
macropulses had a repetition rate of 5~Hz. The extraction of the
absolute photocurrent magnitude in response to a such short pulses
is not an easy task. To calibrate the data we used $\approx
100$~ns long pulses of a transversely excited atmospheric pressure
(TEA) CO$_2$ laser with a fixed operating wavelength of $\lambda =
9.6$~$\mu$m (corresponding photon energy $\hbar \omega = 129$~meV)
and power $P \simeq $~2~kW. The wavelength dependence of the
current from FELIX was scaled to coincide with the CO$_2$ laser
result. For the measurements in the terahertz range we used a
molecular gas laser, which was pumped optically by a TEA CO$_2$
laser.~\cite{GanichevPrettl} With  NH$_3$ as the active laser gas,
we obtain 100~ns pulses of linearly polarized radiation  at
wavelengths $\lambda= 90$, 148 and 280~$\mu$m (corresponding
photon energies $\hbar \omega$ are 13.7, 8.4 and 4.4~meV). We also
used D$_2$O and CH$_3$F as laser gases to obtain radiation with
$\lambda= 385$ and 496~$\mu$m ($\hbar \omega = 3.2$ and
$2.5$~meV), respectively. The peak power of THz radiation used in
our experiments was in the range from 3~kW to 30~kW. Radiation was
applied with an angle of incidence $\theta_0$ varying from
$-30^\circ$ to +30$^\circ$ to the QW normal in the ($xz$) plane,
see inset to Fig.~\ref{fig01}. To investigate the circular
photogalvanic effect we use elliptically polarized light. The
polarization of the laser beam was modified from linear to
elliptical (and circular) by means of ZnSe Fresnel rhombus for the
mid-infrared radiation and crystal quartz $\lambda/4$-plates for
the THz radiation. The radiation helicity $P_{\rm circ}$ of the
incident light varies from -1 ($\sigma_{-}$) to +1 ($\sigma_{+}$)
according to $P_{\rm circ} = \sin 2\varphi$, where $\varphi$ is
the angle between the initial plane of the laser radiation
polarization and the optical axis of the quarter-wave polarizer.
The light polarization states for some characteristic angles
$\varphi$ are sketched on the top of Fig.~\ref{fig01}.

\section{Experimental results}
\label{experiment}

Irradiating (001)-grown HgTe/HgCdTe quantum well structures with
polarized light at oblique incidence we detected a photocurrent
signal measured across one of the contact pairs.
Figure~\ref{fig01} (left panel) shows the  dependence of the
photocurrent on the angle $\varphi$ measured at room temperature
in the direction normal to the plane of incidence ($xz$) and
obtained at two angles of incidence $\theta_0 = \pm 30^\circ$. The
current reversed its direction when switching the sign of the
radiation helicity. For the shown data, the polarization
dependence of the photocurrent is well described by $J_y = J_0
\cdot \sin2 \varphi \propto P_{\rm circ}$.~\cite{footnote11} We
emphasize that by inverting the incidence angle the photocurrent
changes its sign (at $\theta_0 \approx 0$). In the longitudinal
geometry, when the photocurrent is picked up in the plane of
incidence, no helicity dependent photocurrent was observed. Below,
in Sec.~\ref{pheno}, we will demonstrate that these polarization
and angle of incidence dependences follow exactly from a
phenomenological theory. The effect is observed in a wide spectral
and temperature range, from 6~$\mu $m to 496~$\mu $m and from
4.2~K to 300~K.

\begin{figure}[t]
\includegraphics[width=0.8\linewidth]{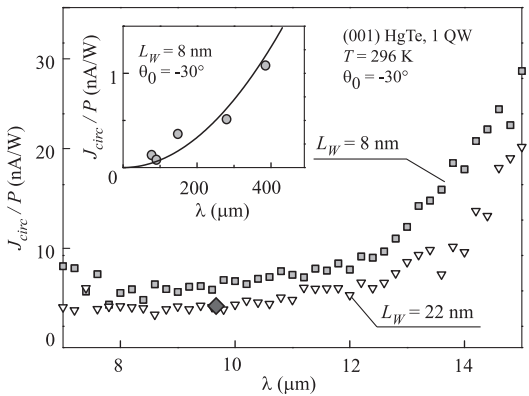}
\caption{The spectral dependence of photon-helicity driven
photocurrent measured in (001)-grown QWs with two QW widths of
$L_W = 22$~nm (triangles) and $L_W= 8$~nm (squares) in the
mid-infrared range. The data were taken using a free electron
laser and were scaled to the CPGE photocurrent obtained for $L_W =
22$~nm found by using a TEA CO$_2$ laser operating at $\lambda =
9.6\, \mu$m (diamond). The inset shows the CPGE photocurrent
spectrum in the THz-range for $L_W = 8$~nm. The full line is a
guide for the eyes. } \label{fig02}
\end{figure}

Figure~\ref{fig02} shows the wavelength dependence of the circular
(photon-helicity-dependent) photocurrent, $J_{\rm circ}$. To
extract $J_{\rm circ}$ from the total signal we used the fact that
it changes direction upon switching the helicity. Taking the
difference of right- and left-handed radiation induced
photocurrents we define $J_{\rm
circ}$\,=\,$[J(\varphi$\,=\,$45^\circ)$\,$-$\,$J(\varphi$\,=\,$135^\circ)]/2$.
Figure~\ref{fig02} demonstrates that in the mid-infrared range
both samples exhibit similar spectral behavior: the signal does
not depend on the wavelength in the range from 6 to 12~$\mu$m and
then rises rapidly, so that from 12 to 15~$\mu$m the photocurrent
strength increases by factor of~4. While the initial lack of
spectral dependence of $J_{\rm circ}$ can be attributed to the
spectral behavior of the interband absorption in this range, the
signal rise at longer wavelength remains unclear. One of the
possible explanations might be attributed to intersubband
resonance absorption. However, the wavelength of resonant
transitions between the two lowest subbands should depend strongly
on the quantum well width, whereas Fig.~\ref{fig02} shows that the
signal increase occurs at the same wavelengths for samples with
very different quantum well widths. This observation excludes
resonant intersubband transitions as an origin of such spectral
behavior, unless there is a accidental coincidence of the
absorption between the lowest subbands in the narrow well and the
absorption into higher subbands in the wider well. This is
possible in principle in HgTe-based QWs which have a complex
 band structure. One can also imagine that, in spite of
the fact that direct optical transitions dominate in the
absorption of mid-infrared radiation, the weaker free carrier
absorption (Drude-like) may contribute substantially to the
photocurrent generation yielding larger signals at lower
frequencies. This assumption, however, is not supported by the
measurements carried out in the THz range, where current is
excited by  radiation with photon energy smaller than the energy
gap and is due to free carrier absorption. The inset in
Fig.~\ref{fig02} demonstrates that, as expected for free carrier
absorption, the signal rises significantly with increasing
wavelength as expected for Drude absorption. We emphasize that in
the whole THz range the photocurrent magnitude is substantially
smaller than that excited by mid-infrared radiation. This fact
shows that the photocurrent in the mid-infrared can not be
attributed to indirect transitions.

\begin{figure}[t]
\includegraphics[width=0.8\linewidth]{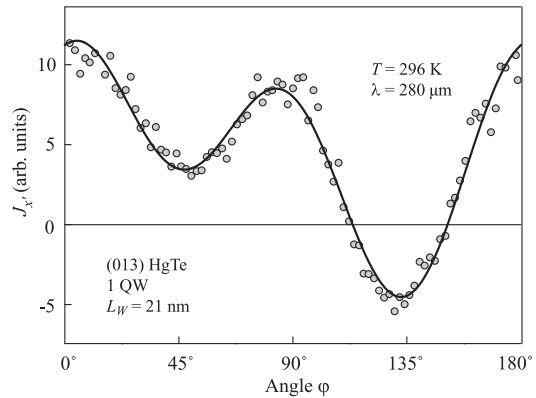}
\caption{The helicity dependence of the photocurrent in
(013)-grown HgTe single QW for normal incidence light. Solid line
is fit to phenomenological Eq.~\protect(\ref{ph_08}).}
\label{fig03}
\end{figure}

Helicity dependent photocurrent was also detected in (013)-grown
samples (see Figs.~\ref{fig01} and~\ref{fig03}). However, in
contrast to (001)-oriented QWs, in these structures the current is
observed even at normal incidence. Depending on the sample type,
radiation wavelength and/or sample temperature the currents in
both $x^\prime$ and  $y^\prime$ directions can be well fitted
simply by $J = J_0 \cdot \sin2 \varphi\propto P_{\rm circ}$ (as
for the circumstances of ~Fig.~\ref{fig01}) or by more complex
dependence on the angle $\varphi$ (see Fig.~\ref{fig03}) given by
\begin{equation}
J=A\cdot \text{sin}2\varphi+B\cdot \text{sin}4\varphi+C\cdot
\text{cos}4\varphi+D \,.
\label{ex_01}
\end{equation}
Here, $A$, $B$, $C$, and $D$ are fitting parameters, described in
more detail below in Sec.~\ref{pheno}. While the first term in the
right-hand side of the Eq.~(\ref{ex_01}) describes the
CPGE,~\cite{Ganichev01} the other terms yield the
LPGE.~\cite{GanichevPrettl,bookIvchenko,SturmanFridkin_book} We
note that even if B, C and D are non-zero, $J_{\rm
circ}$\,=\,$[J(\varphi$\,=\,$45^\circ)$\,$-$\,$J(\varphi$\,=\,$135^\circ)]/2$
is proportional to A and hence is a measure of the CPGE. Measuring
the CPGE and LPGE contributions as a function of the angle of
incidence we observed that all currents for the (013)-grown
samples reach a maximal value at normal incidence. Below we focus
on CPGE.

\begin{figure}[t]
\includegraphics[width=0.8\linewidth]{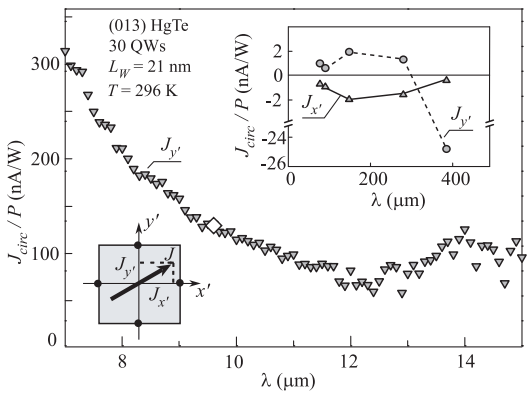}
\caption{The spectral behavior of the power-normalized
helicity-dependent photocurrent along the $y'$ direction measured
in (013)-grown QW excited by the mid-infrared radiation. The data
are obtained at normal incidence using the free electron laser
(down triangles) and normalized to the CPGE photocurrent measured
applying TEA CO$_2$ laser (diamond) operating at $\lambda = 9.6\,
\mu$m. The top right inset shows spectral behavior of
$J_{x^\prime}$ and $J_{y^\prime}$ in the THz-range, which
correspond to the projections of the total photocurrent (see the
inset in the bottom, left). Note the axis break on the ordinate. }
\label{fig04}
\end{figure}

\begin{figure}[t]
\includegraphics[width=0.8\linewidth]{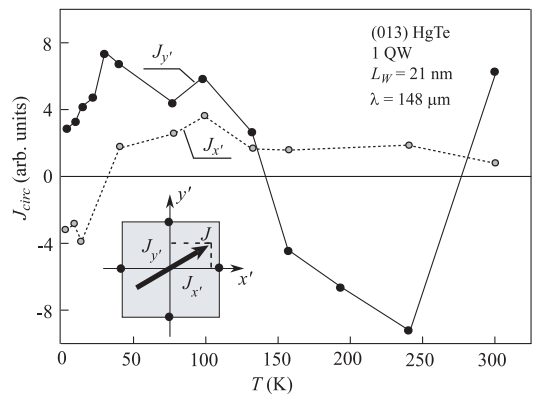}
\caption{Temperature dependence of  $J_{x^\prime}$ and
$J_{y^\prime}$ photocurrents measured in (013)-grown single QW
applying radiation with $\lambda = 148$~$\mu$m. The photocurrents
correspond to the  projections of the total photocurrent on
$x^\prime$ and $y^\prime$ directions (see the inset). }
\label{fig05}
\end{figure}

Experimentally, we observed a major difference between the
photocurrent excited in (013)-oriented QWs and that in (001)-grown
QWs. In the higher symmetry structures, (001)-grown QWs,  $J_{\rm
circ}$ can only be excited by oblique incidence, its direction is
always restricted to the perpendicular to the plane of incidence
for the used experimental geometry, and only the photocurrent
magnitude and sign are functions of angle, temperature and
wavelength.  In the (013)-grown structure, $J_{\rm circ}$ can be
produced by normal incidence and can flow in any direction as
shown in Figs.~\ref{fig01}, \ref{fig04} and~\ref{fig05}, where it
can be seen that the photocurrent components  $x^\prime$ and
$y^\prime$ behave differently and even change their relative sign
depending on the wavelength and temperature. The spectral behavior
of the circular photocurrent is shown in Fig.~\ref{fig04} for one
of the pairs of contacts. As for (001)-grown QWs, we observed that
for some of the mid-infrared spectral range (in this case $10 \div
15~\mu$m) the signal is almost independent of the wavelength.
Unlike (001)-grown QWs, it becomes larger at shorter $\lambda$.
Applying THz radiation we observed the  photocurrent caused by
free carrier absorption which, like for (001)-grown QWs, is
substantially smaller  than that in mid-infrared range. The
wavelength and the temperature dependence of $J_{x^\prime}$ and
$J_{y^\prime}$ components are shown in the inset in
Fig.~\ref{fig04} and in Fig.~\ref{fig05}, respectively. We
emphasize that the lack of restriction on the photocurrent
direction was detected for (013)-grown QWs only, which are
characterized by the point symmetry group C$_1$, the group of the
lowest symmetry comprising only the identity.

\section{Phenomenological theory and discussion}
\label{pheno}

In order to describe the observed  dependences of photocurrent on
the light polarization and the angle of incidence we derive here
phenomenological equations for the  linear and circular
photogalvanic effects in two-dimensional HgTe-based structures.
The photogalvanic current density $\bm{j}$ can be written as a
function of the electric component $\bm{E}$ of the radiation field
and the propagation direction $\hat{\bm{e}}$ in the following
form~\cite{IvchenkoGanichev_book,bookIvchenko}
\begin{equation}
\label{photogalv}
j_{\lambda} =
\sum_{\beta}
\gamma_{\lambda \beta}\:\hat{e}_\beta P_{\rm circ}\:|E|^2 +
\sum_{\mu, \nu}
\chi_{\lambda \mu \nu} \frac{E_{\mu} E^*_{\nu} + E_{\mu}^* E_{\nu}}{2}\:,
\end{equation}
where the first term on the right hand side is proportional to the
radiation helicity $P_{\rm circ}$ and represents the CPGE, while
the second term corresponds to the linear photogalvanic
effect,~\cite{GanichevPrettl,bookIvchenko} which may be
superimposed on the CPGE. The indices $\lambda$, $\beta$, $\mu$,
$\nu$ run over the coordinate axes. The second rank pseudotensor
$\bm{\gamma}$ and the third rank tensor $\bm{\chi}$, symmetric in
the last two indices, are material parameters.  We note that while
in the theoretical consideration the current density $\bm{j}$ is
used, in the experiments the electric current $\bm{J}$ is measured
which is proportional to $\bm{j}$.

\begin{figure}[t]
\includegraphics[width=0.99\linewidth]{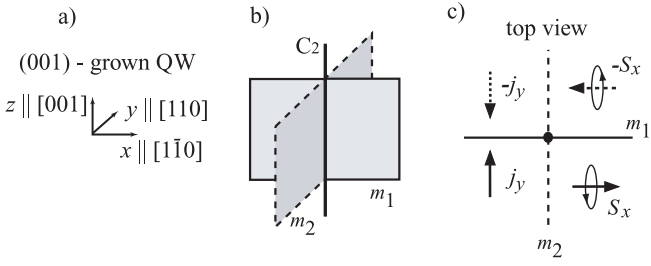}
\caption{(a) Coordinate system of the (001)-grown HgTe QW which
has C$_{2\mathrm{v}}$ symmetry, (b) mirror planes $m_1$ and $m_2$
and C$_2$-axis in QW grown along $z\,
\parallel [001]$. Arrows in the drawing (c) show that the
reflection in the mirror plane $m_1$ changes the sign of both the
polar vector component $j_y$ and the axial vector component $S_x =
P_{\rm{circ}} {\hat e}_x$, demonstrating that a linear coupling
$j_y \propto S_x$ is allowed under these symmetry operations. This
coupling is also allowed by the other symmetry operations of the
point group, and so is $j_x  \propto  S_y$. } \label{fig06}
\end{figure}

The HgTe QWs grown along $z  \parallel [001]$ direction correspond
to the C$_{{\rm 2v}}$ point group. This group includes a rotating
axis C$_2$ parallel to the $\left[ 001 \right]$-direction and two
mirror planes, $m_1$ and $m_2$,  coinciding with the ($xz$) and
($yz$) planes, respectively (Fig.~\ref{fig06}). It follows from
Neumann's Principle and Eq.~(\ref{photogalv}) that the circular
photocurrent can only occur along those axes where for all
symmetry operations components of $\bm{j}$ transform in the same
way as components of the pseudovector $\bm{S} = P_{\rm{circ}}
\hat{\bm{e}}$ describing the radiation helicity. Let us consider
it for circularly polarized radiation propagating along
$x$-direction, i.e., for $S_x$. The reflection in each mirror
plane transforms the current component $j_y$ and the pseudovector
component $S_x$ in the same way: $j_y \rightarrow -j_y$, $S_{x}
\rightarrow -S_{x}$ for the plane $m_1$ (see Fig.~\ref{fig06}c)
and $j_y \rightarrow j_y$, $S_{x} \rightarrow S_{x}$ for the plane
$m_2$. Therefore, the photocurrent $j_y \propto S_x$ is possible.
Similar arguments hold for $j_x \propto S_{y}$.
For any other relative directions of  $\bm{j}$ and $\bm{S}$, a
linear coupling of the current and the radiation helicity is
forbidden. For instance, the reflections in both $m_1$ and $m_2$
planes reverse the direction of $S_z$ while C$_2$ does nothing.
There is no polar vector component that transforms in the same
way. It indicates that $S_z$ cannot give rise to a photocurrent in
(001)-grown QW structures. Thus, CPGE can only be generated at
oblique incidence and in the direction normal to the plane of
incidence, ($xz$) or ($yz$), as observed in the experiments. For
QW structures of the C$_{{\rm 2v}}$ point group, these arguments
can be used to show that the non-zero components of the second
rank pseudotensor and the third rank tensor are the following:
$\gamma_{xy}$, $\gamma_{yx}$, $\chi_{xxz}=\chi_{xzx}$, and
$\chi_{yyz}=\chi_{yzy}$. These four components are linearly
independent. Optical excitation of such structures at oblique
incidence in the $(xz)$-plane with elliptically polarized light
generates electric current whose $x$- and $y$-components depend on
the angles $\varphi$ and $\theta$ as follows
\begin{eqnarray}
 j_x (\varphi) &=&  \frac12 \chi_{xxz} E_0^2 \,t_p^2 \sin\theta \cos\theta \, (1-\cos 4 \varphi) \:, \label{ph_05} \\
 j_y (\varphi) &=&  E_0^2 \, t_p t_s \sin\theta \, \left[ \gamma_{yx}  \sin 2\varphi - \frac12  \chi_{yyz} \sin 4 \varphi\right].
 \label{ph_06}
\end{eqnarray}
Here,  $E_0$ is the electric field amplitude of the incident radiation,
$t_p$ and $t_s$ are the Fresnel amplitude transmission coefficients from vacuum to the
structure for the $s$- and $p$-polarized light,
respectively,~\cite{BornWolf} $\theta$ is the refraction angle related to the incidence
angle $\theta_0$ by $\sin\theta= \sin\theta_0/n_{\omega}$, and $n_{\omega}$
is the refractive index.

We note that the linear photogalvanic effect described by
Eq.~(\ref{ph_05}) and second term on the right hand side of
Eq.~(\ref{ph_06}) was detected for several wavelengths. It can be
excited by linearly polarized radiation, which was also checked
experimentally (the data are not shown here).

The (013)-oriented QWs belong to the trivial point group C$_1$
lacking any symmetry operation except the identity. Hence,
symmetry does not impose any restriction on the relation between
radiation electric field and photocurrent components. All
components of the pseudotensor $\bm{\gamma}$ and the tensor
$\bm{\chi}$ may be different from zero. Phenomenologically, for
the C$_1$-symmetry group, the lateral photogalvanic current for
the excitation along the QW normal with elliptically polarized
light is given by
\begin{eqnarray}
 j_{x'} &=& - E_0^2 \,t_s^2 \left[ \gamma_{x'z'} \sin 2\varphi - \frac{\chi_{x'x'x'}+\chi_{x'y'y'}}{2} \right.  \label{ph_07} \\
       &+& \left. \frac{\chi_{x'x'x'}-\chi_{x'y'y'}}{4} (1+\cos4\varphi) +\frac{\chi_{x'x'y'}}{2}\sin4\varphi \right] \nonumber \:, \\
 j_{y'} &=& - E_0^2 \,t_s^2 \left[ \gamma_{y'z'} \sin 2\varphi - \frac{\chi_{y'x'x'}+\chi_{y'y'y'}}{2} \right.  \label{ph_08} \\
       &+& \left. \frac{\chi_{y'x'x'}-\chi_{y'y'y'}}{4} (1+\cos4\varphi) +\frac{\chi_{y'x'y'}}{2}\sin4\varphi \right]  \:. \nonumber
\end{eqnarray}
Exactly this polarization dependence is encapsulated in
Eq.~(\ref{ex_01}) and observed in experiment as can be seen in
Fig.~\ref{fig01} and, in particular, in Fig.~\ref{fig03}, where
all terms given by Eq.~(\ref{ph_08}) contribute substantially.
Equations~(\ref{ph_07}) and~(\ref{ph_08}) show that, in
(013)-oriented QWs, the CPGE photocurrent direction is arbitrary
and not forced to a definite crystallographic axis.

The fact that $J_{x'}$ and $J_{y'}$ as well as their ratio
$J_{x'}/ J_{y'}$ exhibit a non-trivial variation with the
radiation wavelength (Fig.~\ref{fig04}) or sample temperature
(Fig.~\ref{fig05}) is also in agreement with Eqs.~(\ref{ph_07})
and~(\ref{ph_08}) yielding for circularly polarized light
$j_{x^\prime}/j_{y^\prime} =  \gamma_{x^\prime z^\prime}/
\gamma_{y^\prime z^\prime}$. Indeed all the components of the
tensor $\bm{\gamma}$ are linearly independent in structures of the
C$_1$ point-group symmetry, so the ratio between them can change
by varying the experimental conditions. There is therefore no
preferential direction of the circular photocurrent forced by the
symmetry arguments. The same is valid, and observed
experimentally, for the linear photogalvanic photocurrent.
Symmetry analysis of Eq.~(\ref{photogalv}) for QWs of the C$_1$
point group shows that $dc$ current excited by linearly polarized
light in the geometry of normal incidence is described by
\begin{eqnarray}
 j_{x'} &=& - E_0^2 \,t_s^2 \left[ \chi_{x'x'y'} \sin 2\alpha - \frac{\chi_{x'x'x'}+\chi_{x'y'y'}}{2} \right. \nonumber  \\
       & & + \left. \frac{\chi_{x'x'x'}-\chi_{x'y'y'}}{2} \cos 2 \alpha \right]  \:, \\
  j_{y'} &=& - E_0^2 \,t_s^2 \left[ \chi_{y'x'y'} \sin 2\alpha - \frac{\chi_{y'x'x'}+\chi_{y'y'y'}}{2} \right. \nonumber  \\
       & & + \left. \frac{\chi_{y'x'x'}-\chi_{y'y'y'}}{2} \cos 2 \alpha \right]  \:,
\end{eqnarray}
where $\alpha$ is the angle between the light polarization plane
and the $x'$ axis. We have checked experimentally (not shown) that
the photocurrent induced under  linearly polarized excitation is
well fitted by these equations.

\section{Band structure and optical transitions}
\label{bandoptical}

\begin{figure}
\includegraphics[width=0.99\linewidth]{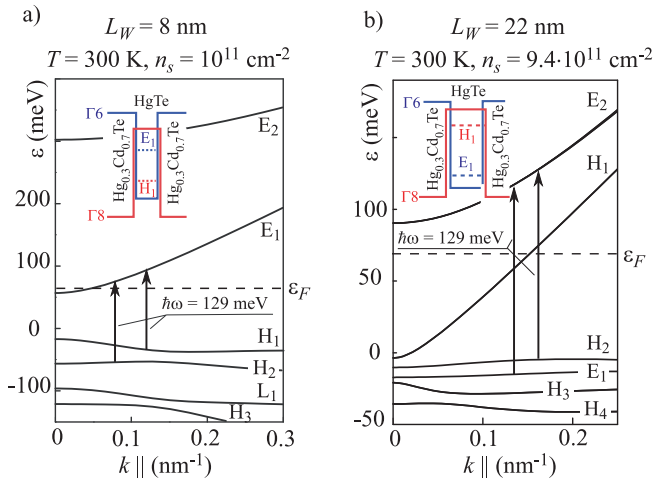}
\caption{Calculated band structure for QWs with $L_W = 8$~nm and
22~nm at $T = 300$~K. Arrows show optical transitions induced by
mid-infrared radiation used in the experiments ($\hbar \omega =
129$~meV). Insets sketch the band profile of HgTe-based QWs with
normal and inverted band structure. } \label{fig07}
\end{figure}

HgTe, which is a semimetal, forms type-III QW with
Hg$_{\scriptstyle 0.3}$Cd$_{\scriptstyle 0.7}$Te barrier as
indicated in the cartoon insets of Fig.~\ref{fig07}. Depending on
the quantum confinement, i.e. the well width, and on the
temperature, a normal (Fig.~\ref{fig07}a) or inverted
(Fig.~\ref{fig07}b) band structure can be realized with positive
or negative gap between   $E_1$- and $H_1$-subbands (see
Ref.~\onlinecite{Koenig2008}).

We have calculated the band structure of HgTe/Hg$_{\scriptstyle
0.3}$Cd$_{\scriptstyle 0.7}$Te QWs grown along [001] direction
using an envelope function approximation~\cite{Burt1999} based on
eight-band $\bm{k} \cdot \bm{p}$ Hamiltonian. The influence of the
induced free carriers has been included by self-consistent
solution of the Poisson and Schr\"{o}dinger equations. The total
single-particle wave function is given by:
\begin{equation}\label{Psi}
  \Psi(\bm{r})=\sum_{n}F_{n}(\bm{r})
  |u_{n0}(\bm{r})\rangle=\sum_{n}\exp^{i\bm{k}_{\|}\bm{r}_{\|}}f_{n}(z)
  |u_{n0}(\bm{r})\rangle,
\end{equation}
where $F_{n}$ are the envelope functions; $|u_{n0}\rangle$ are the
Bloch functions at the $\Gamma$ point of the Brillouin zone, which
form a complete basis set, and are assumed to be the same in HgTe-
and Hg$_{\scriptstyle 0.3}$Cd$_{\scriptstyle 0.7}$Te-layers;
$\bm{k}_{\|}$ is the wave vector in the plane of the QW. For
narrow gap structures it is important to include the coupling
between the conduction and valence bands, and the spin-orbit
coupling of the valence bands. Therefore, the conduction and
valence  Bloch functions ($|\Gamma_6,\pm1/2\rangle$)
($|\Gamma_8,\pm1/2\rangle$, $|\Gamma_8,\pm3/2\rangle$,
$|\Gamma_7,\pm1/2\rangle$) were used as a basis
set.~\cite{Kane1957} In this case the following system of eight
coupled differential equations has to be solved in order to find
the envelope functions and the energies:
\begin{equation}\label{Eignprob}
  \sum_{n'} \left[ \sum_{\alpha}P_{nn'}^{\alpha}k_{\alpha}+\sum_{\alpha,\beta}
k_{\alpha}D_{nn'}^{\alpha\beta}k_{\beta} +H^{BP}_{nn'} \right] f_{n'}(z)
\end{equation}
\[
+ [ E_{n}(z) + V(z) ] f_{n}(z)=E\cdot f_{n}(z) \:.
\]
Here,  indices $n$ and $n'$ run over the two conduction and six
valence bands, $E_{n}(z)$ is the respective band-edge potential;
indices $\alpha$ and $\beta$ run over $x$, $y$, $z$; $k_x$ and
$k_y$ are components of the in-plane wave vector, $k_{z}$ is the
operator $-i
\partial/\partial z$; $P_{nn'}^{\alpha}$ is the momentum matrix element describing the
interaction between two bands of the chosen basis set
$|u_{n0}\rangle$,  $D_{nn'}^{\alpha\beta}$ describes the coupling
between the bands in this basis set and remote bands in the
second-order perturbation theory, $V(z)$ is the self-consistently
calculated Hartree-potential, and $H^{BP}_{nn'}$ are the
strain-induced terms of the Bir-Pikus Hamiltonian.~\cite{Bir74} A
detailed description of the model as well as the band structure
parameters employed in the calculations can be found in
Ref.~\onlinecite{Novik2005}.
Our calculations show that mid-infrared radiation with a photon energy of the order of
100~meV used in the experiments causes direct
interband optical transitions in all our samples (see Fig.~\ref{fig07}).

\section{Conclusions}
\label{summary}

To summarize, our experiments show that helicity-driven
photogalvanic currents can effectively be generated in HgTe
quantum wells. The photocurrent strength, e.g., for CPGE current,
of about an order of magnitude larger than that observed in GaAs,
InAs and SiGe low dimensional
structures.\cite{IvchenkoGanichev_book} Our results reveal that
photogalvanic measurements open a rich field for investigation of
microscopic properties of this novel and promising material.
The microscopic theory of these phenomena in HgTe-based QWs, in
particular with the inverted bands ordering, is a task for future,
which is complicated by the band structure, with almost flat
valence bands and strong spin slitting in comparison with other
materials. All known mechanisms of the CPGE are enhanced in
HgTe-based structures: spin-related
mechanisms~\cite{IvchenkoGanichev_book,Ganichev01,GanichevPrettlreview,Bieler05}
due to large spin-orbit coupling and orbital
mechanisms~\cite{Tarasenko07,Olbrich09} due to very small band
gap.  We can suggest that the photocurrent induced by the
free-carrier absorption of THz radiation is dominated by orbital
mechanisms. For the direct optical transitions, spin-related
mechanisms of the current formation can play an important role. To
distinguish between spin and orbital mechanism additional research
is required. In this respect future experiments on time resolved
photogalvanics under short-pulsed circularly polarized
photoexcitation, with the pulse duration being comparable with the
free-carrier momentum and spin relaxation times, would be
desirable and informative. Such experiments would also reveal a
great deal about the momentum, energy and spin relaxation of
nonequilibrium photoexcited carriers. The experiments will also
clarify the importance of the recently proposed Berry-phase based
effects which should be enlarged in narrow gap materials.

\section*{ACKNOWLEDGMENTS}
We thank E.L.\,Ivchenko for helpful discussions. The financial
support of the DFG via programs WE2476/9-1 and AS327/2-1, the
Linkage Grant of IB of BMBF at DLR, the RFBR,  the President Grant
for young scientists (MD-1717.2009.2), and Foundation
"Dynasty"-ICFPM is gratefully acknowledged.

* (for NQVinh) Present address ITST, Department of Physics, University
of California, Santa Barbara CA 93106-4170.


\begin{thebibliography}{99}
%
\bibitem{Buhmann09} H. Buhmann, Int. J. Mod. Phys. B  \textbf{23},  2551 (2009).

\bibitem{Koenig07} M. K\"{o}nig, S. Wiedmann, C. Br\"{u}ne, A. Roth, H. Buhmann, L.W. Molenkamp, X.-L. Qi,
and S.-C. Zhang, Science \textbf{318}, 766 (2007).

\bibitem{Gui04} Y.S.~Gui, C.R.~Becker, N.~Dai, J.~Liu, C.J.~Qui, E.G.~Novik,
M.~Sch\"afer, X.Z.~Shu, H.J.~Chu, H.~Buhmann, and L.W.~Molenkamp,
Phys. Rev. B {\bf 70}, 115328 (2004).

\bibitem{Bernevig06} B.A. Bernevig \textit{et al.}, Science \textbf{314}, 1757 (2006).

\bibitem{Diehl09} H.~Diehl,  V.A.~Shalygin, L.E.~Golub, S.A.~Tarasenko, S.N.~Danilov,
V.V.~Bel'kov, E.G.~Novik, H.~Buhmann, L.W. Molenkamp,
C.~Br\"{u}ne, E.L.~Ivchenko, and S.D.~Ganichev, Phys. Rev. B
\textbf{80}, 075311 (2009).

\bibitem{IvchenkoGanichev_book} E.L. Ivchenko and S.D.~Ganichev,
\textit{Spin Photogalvanics} in \textit{Spin Physics in Semiconductors},
ed. M.I.~Dyakonov, (Springer, Berlin, 2008).

\bibitem{Ganichev01} S.D.~Ganichev, E.L.~Ivchenko, S.N.~Danilov,
J.~Eroms, W.~Wegscheider, D.~Weiss, and W.~Prettl, Phys. Rev.
Lett. {\bf 86}, 4358 (2001).

\bibitem{GanichevPrettlreview} S.D.~Ganichev and  W.~Prettl,
J. Phys.: Condens. Matter, {\bf 15}, R935 (2003).

\bibitem{Bieler05} M.~Bieler, N.~Laman, H.M.~van~Driel, and
A.L.~Smirl, Appl. Phys. Lett. {\bf 86}, 061102 (2005).

\bibitem{Tarasenko07} S.A.\,Tarasenko,
JETP Lett. {\bf 85}, 182 (2007).

\bibitem{Olbrich09} P. Olbrich, S.A. Tarasenko, C. Reitmaier, J. Karch, D. Plohmann, Z.D. Kvon, and S.D. Ganichev,
Phys. Rev. B \textbf{79}, 121302(R) (2009).

\bibitem{Deyo09} E. Deyo, L.E. Golub, E.L. Ivchenko, and B.~Spivak, arXiv:0904.1917 (2009).

\bibitem{Moore09} J.E. Moore  and J. Orenstein,
arXiv:0911.3630 (2009).

\bibitem{JAP2008}  S.D. Ganichev, W. Weber, J. Kiermaier, S.N.~Danilov,  P.~Olbrich, D.~Schuh,
W.~Wegscheider, D. Bougeard, G. Abstreiter, and W.~Prettl, J. Appl. Phys. \textbf{103}, 114504 (2008).

\bibitem{MBE} V.S. Varavin \textit{et al.}, Proc. SPIE \textbf{381}, 5136 (2003).

\bibitem{Knippels99p1578}  G.\,M.\,H. Knippels, X. Yan, A.\,M. MacLeod,
W.\,A. Gillespie, M. Yasumoto, D. Oepts, and A.\,F.\,G. van der
Meer, Phys. Rev. Lett. \textbf{ 83}, 1578 (1999).

\bibitem{GanichevPrettl}
S.D. Ganichev and W.~Prettl, \textit{Intense Terahertz Excitation of
Semiconductors} (Oxford Univ. Press, 2006).

\bibitem{footnote11} We note that at some wavelength of the radiation an addition
photocurrent component proportional to $\sin 4 \varphi$ is
detected. This contribution is due to linear photogalvanic
effect.~\protect
\cite{GanichevPrettl,bookIvchenko,SturmanFridkin_book}

\bibitem{bookIvchenko} E.L. Ivchenko, \textit{Optical Spectroscopy of
Semiconductor Nanostructures} (Alpha Science International,
Harrow, UK, 2005).

\bibitem{SturmanFridkin_book} B.I.~Sturman and V.M.~Fridkin, \textit{The
Photovoltaic and Photorefractive Effects in Non-Centrosymmetric
Materials} (Nauka, Moscow, 1992; Gordon and Breach, New York,
1992).

\bibitem{BornWolf} M.~Born and  E.~Wolf, \textit{Principles of Optics} (Pergamon Press, Oxford, 1970).

\bibitem{Koenig2008} M. K\"{o}nig, H. Buhmann, L.W. Molenkamp, T. Hughes, C.-X. Liu,
X.-L. Qi, and S.-C. Zhang,  J. Phys. Soc. Jpn. \textbf{77}, 031007 (2008).

\bibitem{Burt1999} M.G. Burt, J. Phys.: Condens. Matter \textbf{4}, 6651 (1992);
\textit{ibid.} \textbf{11}, R53 (1999); B.A. Foreman, Phys. Rev. B \textbf{48}, 4964 (1993).

\bibitem{Kane1957}
E.O. Kane, J. Phys. Chem. Solids \textbf{1}, 249 (1957).

\bibitem{Bir74}
G.L.~Bir and G.E.~Pikus, \textit{Symmetry and Strain-Induced
Effects in Semiconductors} (John Wiley {\&} Sons, Chichester,
1974).

\bibitem{Novik2005}
E.G. Novik, A. Pfeuffer-Jeschke, T. Jungwirth, V. Latussek, C.R. Becker,
G. Landwehr, H. Buhmann, and L.W. Molenkamp,
Phys. Rev. B \textbf{72}, 035321 (2005).

\end{thebibliography}
\end{document}